\documentclass[reprint,showpacs,preprintnumbers,pre,superscriptaddress]{revtex4-2}

 \usepackage{graphicx}
 \usepackage{bm}
 \usepackage[colorlinks,linkcolor=blue,anchorcolor=blue,citecolor=blue,urlcolor=blue]{hyperref}
 \usepackage{amsmath}

 \newcommand{\e}{\mathrm{e}}
 \newcommand{\C}{\mathrm{C}}
 \newcommand{\dif}[1]{\mathrm{d}#1\mathop{}\!}

\begin{document}

    \title{Microscopic low-dissipation heat engine via shortcuts to adiabaticity and shortcuts to isothermality}

    \author{Xiu-Hua Zhao}
    \affiliation{Department of Physics, Beijing Normal University, Beijing 100875, China}
    \author{Zheng-Nan Gong}
    \affiliation{Zun Yi Si Zhong, Zunyi 563000, China}
    \author{Z. C. Tu}\email[Corresponding author. Email: ]{tuzc@bnu.edu.cn}
    \affiliation{Department of Physics, Beijing Normal University, Beijing 100875, China}

    \date{\today}

    \begin{abstract}
        We construct a microscopic model of low-dissipation engines by driving a Brownian particle in a time-dependent harmonic potential. Shortcuts to adiabaticity and shortcuts to isothermality are introduced to realize the adiabatic and isothermal branches in a thermodynamic cycle, respectively. We derive an analytical expression of the efficiency at maximum power for this kind of engines. This expression satisfies the universal law of efficiency at maximum power up to the second order of the Carnot efficiency. We also analyze the issue of power at any given efficiency for general low-dissipation engines, and then obtain the supremum of the power in three limiting cases respectively.
    \end{abstract}
    
    \maketitle

    \section{\label{Sec:I}INTRODUCTION}
    Finite-time thermodynamics~\cite{chenAdvancesFiniteTime2004a,bejanAdvancedEngineeringThermodynamics2016a} is a new active branch of nonequilibrium physics. One of the most important topics in finite-time thermodynamics is the efficiency at maximum power for heat engines. Researchers have investigated the efficiencies at maximum power for various models of finite-time heat engines, including endoreversible Carnot-like engine~\cite{curzonEfficiencyCarnotEngine1975,chenEffectHeatTransfer1989}, stochastic engine~\cite{t.schmiedlEfficiencyMaximumPower2007}, Feynman's ratchet~\cite{tuEfficiencyMaximumPower2008}, quantum dot engine~\cite{espositoThermoelectricEfficiencyMaximum2009}, low-dissipation engine~\cite{espositoEfficiencyMaximumPower2010}, minimally nonlinear irreversible engine~\cite{izumidaEfficiencyMaximumPower2012} and so on. Their studies reveal an impressive universality that under certain conditions the efficiencies at maximum power for different models are identical up to the quadratic term of Carnot efficiency~\cite{curzonEfficiencyCarnotEngine1975,chenEffectHeatTransfer1989,t.schmiedlEfficiencyMaximumPower2007,tuEfficiencyMaximumPower2008,espositoThermoelectricEfficiencyMaximum2009}. It is found that the universality up to linear term is due to the tight coupling~\cite{vandenbroeckThermodynamicEfficiencyMaximum2005}, and that the universality up to quadratic term owes to symmetric coupling~\cite{espositoUniversalityEfficiencyMaximum2009} or energy matching~\cite{shengConstitutiveRelationNonlinear2015,tuAbstractModelsHeat2021}. 

    To seek both powerful and efficient engines for practical applications, an increasing number of researchers have been devoting themselves to study the general constraints for efficiency and power~\cite{chenEffectHeatTransfer1989,gordonGeneralPerformanceCharacteristics1992a,chenCurzonAhlbornEfficiency2001,espositoEfficiencyMaximumPower2010,wangBoundsEfficiencyMaximum2012a,izumidaEfficiencyMaximumPower2012,wangBoundsEfficiencyMaximum2013,holubecEfficiencyMaximumPower2015,holubecMaximumEfficiencyLowdissipation2016,ryabovMaximumEfficiencySteadystate2016,naotoshiraishiUniversalTradeOffRelation2016,dechantUnderdampedStochasticHeat2017b,pietzonkaUniversalTradeOffPower2018,maUniversalConstraintEfficiency2018a}. For endoreversible engines, Chen and Yan derived an optimum relation between power and efficiency~\cite{chenEffectHeatTransfer1989} while Gordon and Huleihil provided the power-versus-efficiency diagram~\cite{gordonGeneralPerformanceCharacteristics1992a}. More recently, Esposito {\it et al}. derived the upper and lower bounds of efficiency at maximum power for low-dissipation engines~\cite{espositoEfficiencyMaximumPower2010}. Holubec and Ryabov~\cite{holubecEfficiencyMaximumPower2015,holubecMaximumEfficiencyLowdissipation2016} discussed the efficiency at arbitrary power for low-dissipation engines and obtained analytical upper bounds of efficiency in the regions nearby the maximum power and the zero power. Ma {\it et al}. analytically derived the constraints on efficiency for all power values~\cite{maUniversalConstraintEfficiency2018a}. But their results are not the supremum or infimum in the whole region since the bounds of efficiency for some power values are inaccessible.

    With the fast development of optical-trap technique, the design and realization of microscopic engines have been widely discussed~\cite{ranaSingleparticleStochasticHeat2014,tuStochasticHeatEngine2014a,holubecExactlySolvableModel2014,blickleRealizationMicrometresizedStochastic2012,martinezBrownianCarnotEngine2016,a.martinezColloidalHeatEngines2017}. Schmiedl and Seifert constructed a stochastic Carnot-like engine by using a time-dependent potential to drive a Brownian particle~\cite{t.schmiedlEfficiencyMaximumPower2007}. The protocol during the isothermal processes is chosen to yield a maximum work output while the adiabatic transitions are completed instantaneously. Considering that the mismatch of kinetic energy in the instantaneously adiabatic transition inevitably results in heat exchange between two heat baths, the last author in the present work proposed replacing the instantaneously adiabatic transitions with shortcuts to adiabaticity~\cite{tuStochasticHeatEngine2014a}. However, the isothermal transitions in these models are not really isothermal in the traditional sense since the effective temperature is time-dependent. This shortage inspires the subsequent researches. Following the work by Salazar and Lira~\cite{salazarStochasticThermodynamicsNonharmonic2019}, Chen {\it et al}. realized the isothermal processes with exponential protocols under the assumption of slow driving~\cite{chenMicroscopicTheoryCurzonAhlborn2021}. Nakamura {\it et al}. developed the fast-forward approach to mimic the finite-time isothermal processes~\cite{nakamuraFastforwardApproachStochastic2020}. Their approach consists of two steps: determining the driving potential in an extremely slow time evolution and then rescaling the time variable so that the Kramers equation works for finite-time regions. A more straightforward approach is the shortcut to isothermality~\cite{liShortcutsIsothermalityNonequilibrium2017,liEquilibriumFreeenergyDifferences2021} which is the correspondence of quasi-static isothermal process in finite-time thermodynamics. The calculations of work and heat are tractable in shortcuts to isothermality. There is still blank in the study of heat engines with the consideration of shortcuts to isothermality.
    
    In this work, we employ both shortcuts to isothermality~\cite{liShortcutsIsothermalityNonequilibrium2017} and shortcuts to adiabaticity~\cite{chenFastOptimalFrictionless2010,jarzynskiGeneratingShortcutsAdiabaticity2013} to accomplish a microscopic Carnot-like engine. This engine turns out to be of low dissipation. We also investigate the power and efficiency of this low-dissipation engine. The rest of this paper is organized as follows. In Sec.~\ref{Sec:II}, we revisit the shortcuts to isothermality and shortcuts to adiabaticity. In Sec.~\ref{Sec:III}, we construct a microscopic heat engine including two isothermal branches and two adiabatic branches. The work dissipated during the isothermal branches is inversely proportional to the operation time, which means what we construct is exactly a model of low-dissipation engines. In Sec.~\ref{Sec:IV}, we calculate the efficiency at maximum power of this engine and find that it satisfies the universal law when the damping coefficients in both isothermal branches are identical. In Sec.~\ref{Sec:V}, we discuss the power at any given efficiency for general low-dissipation engines and obtain the analytical supremum of power when the ratio of dissipation coefficients during the cold and hot isothermal branches approaches zero, one and infinity, respectively. In Sec.~\ref{Sec:VI}, we compare our results for constraints on efficiency at given power with those in Refs.~\cite{holubecMaximumEfficiencyLowdissipation2016,maUniversalConstraintEfficiency2018a}. The last section is a brief summary. 
    
    \section{\label{Sec:II}revisiting shortcuts to isothermality and shortcuts to adiabaticity}
    In this section, we outline two key concepts that we will adopt in this work. The first one is the shortcut to isothermality and the second one is the shortcut to adiabaticity. 

    \subsection{\label{Sec:IIA}Shortcuts to isothermality}
    Consider a Brownian particle moving in a one-dimensional potential $U_o(x,\lambda)$, where $\lambda=\lambda(t)$ is a time-dependent external parameter. The Hamiltonian of this particle is $H_o=p^2/2+U_o(x,\lambda)$. The mass of the particle is set to be unit for convenience. The Brownian particle is coupled to a heat bath with constant temperature $T$. To realize finite-time isothermal transitions between two equilibrium states with same temperature, Li {\it et al}.~\cite{liShortcutsIsothermalityNonequilibrium2017,liEquilibriumFreeenergyDifferences2021} proposed a framework of shortcuts to isothermality by introducing an auxiliary potential $U_a(x,p,t)$ so that the distribution function of the system always maintains the following canonical form
    \begin{equation}
        \label{ieq}
            \rho=\e^{\beta F(\lambda)-\beta H_o(x,p,\lambda)},
    \end{equation}
    where $F=-\beta^{-1}\ln\left[\int\int\dif{x}\dif{p}\e^{-\beta H_o(x,p,\lambda)} \right]$ and $\beta=1/T$. We have set the Boltzmann constant to be unit. The auxiliary potential can be determined by substituting Eq.~(\ref{ieq}) to the generalized Kramers equation (see Eq.~(19) in Ref.~\cite{liShortcutsIsothermalityNonequilibrium2017}). As for the time-dependent harmonic potential $U_o=\lambda^2(t)x^2/2$, the auxiliary potential is 
    \begin{equation}
        U_a=\frac{\dot{\lambda}(t)}{2\gamma\lambda(t)}\left[(p-\gamma x)^2+\lambda^2(t)x^2\right],
    \end{equation}
    where $\gamma$ is the damping coefficient and the dot above a character denotes the derivative with respect to time $t$. 
    To ensure that the initial and finial states of the system are in equilibrium with the bath, a constraint
    \begin{equation}
        \label{bc}
        \dot{\lambda}(t_i)=\dot{\lambda}(t_f)=0
    \end{equation}
    should be imposed at the initial time $t_i$ and the finial time $t_f$~\cite{liShortcutsIsothermalityNonequilibrium2017}. 

    According to the stochastic thermodynamics~\cite{jarzynskiNonequilibriumEqualityFree1997,sekimotoStochasticEnergetics2010a,seifertStochasticThermodynamicsFluctuation2012}, the ensemble-averaged work exerted on the particle during the shortcut to isothermality is
    \begin{equation}
        \label{Wiso}
        \begin{aligned}
            W&\equiv\left\langle\int_{t_i}^{t_f}\dif{t} \frac{\partial H}{\partial t}\right\rangle\\
            &=\int_{t_i}^{t_f}\dif{t}\int\int\dif{x}\dif{p}\rho\left(\frac{\partial U_o}{\partial t}+\frac{\partial U_a}{\partial t}\right)\\
            &=T\ln\frac{\lambda(t_f)}{\lambda(t_i)}+T\frac{C[\Lambda(\tilde{t})]}{t_f-t_i},
        \end{aligned}
    \end{equation}
    where 
    \begin{equation}
        \label{C-func}
        C[\Lambda(\tilde{t})]\equiv\int_0^1\dif{\tilde{t}}\left(\frac{1}{\gamma}+\frac{\gamma}{\Lambda^2}\right)\frac{1}{\Lambda^2}\left(\frac{\dif{\Lambda}}{\dif{\tilde{t}}}\right)^2,
    \end{equation}
    and $\Lambda(\tilde{t})\equiv\lambda(t_i+(t_f-t_i)\tilde{t})$. $\left\langle\dots\right\rangle$ denotes the ensemble average under canonical distribution.
    The work in Eq.~(\ref{Wiso}) is decomposed into two parts. The first term equals to the variation of the free energy. The second term represents the dissipative work which is inversely proportional to the operation time $(t_f-t_i)$. Although this result is based on the harmonic potential, the inverse-proportion relation between the dissipative work and the operation time is universal for shortcuts to isothermality~\cite{liShortcutsIsothermalityNonequilibrium2017}. 

    \subsection{\label{Sec:IIB}Shortcuts to adiabaticity}
    Shortcuts to adiabaticity are strategies devised to circumvent the condition of infinitely slow evolution in quantum adiabatic theorem~\cite{emmanouilidouSteeringEigenstateDestination2000,demirplakAdiabaticPopulationTransfer2003,berryTransitionlessQuantumDriving2009,chenFastOptimalFrictionless2010,delcampoShortcutsAdiabaticityCounterdiabatic2013,dengBoostingWorkCharacteristics2013,campoMoreBangYour2014,deffnerClassicalQuantumShortcuts2014,guery-odelinShortcutsAdiabaticityConcepts2019}. These strategies are also applicable to classical systems. In classical mechanics, the volume $\Omega$ of phase space enclosed by an energy shell remains constant when the external parameter varies slowly enough. Jarzynski introduced a counterdiabatic driving Hamiltonian $H_c$ so that $\Omega$ will keep constant even if the external parameter changes at finite rate~\cite{jarzynskiGeneratingShortcutsAdiabaticity2013}. As an example, Jarzynski obtained $H_c=-\dot{\lambda}(t)xp/(2\lambda(t))$ for $H_o=p^2/2+\lambda^2(t)x^2/2$. Hence, the total Hamiltonian becomes
    \begin{equation}
        \label{adiH}
        H\equiv H_o+H_c=\frac{p^2}{2}+\frac{1}{2}\lambda^2(t)x^2-\frac{\dot{\lambda}(t)}{2\lambda(t)}x p.
    \end{equation}
    
    Consider the system described by the above equation decoupling from heat bath and evolving from an equilibrium state with temperature $T_i$ at the initial time $t_i$ to another equilibrium state with temperature $T_f$ at the final time $t_f$. In Ref.~\cite{tuStochasticHeatEngine2014a}, the author found that the shortcuts to adiabaticity can be adopted to accomplish the evolution along an adiabatic path as long as the external parameter satisfies
    \begin{equation}
        \label{adipro}
        \frac{\lambda(t_i)}{T_i}=\frac{\lambda(t_f)}{T_f}.
    \end{equation}
    Moreover, $\dot{\lambda}(t)$ should satisfies the same constraint as Eq.~(\ref{bc}) such that $H=H_o$ at the initial and final time. The adiabatic transition based on shortcuts to adiabaticity can be completed arbitrarily fast.

    \section{\label{Sec:III}MODEL AND ENERGETICS}
\begin{figure}[!htpb]
        \includegraphics[width=0.4\textwidth]{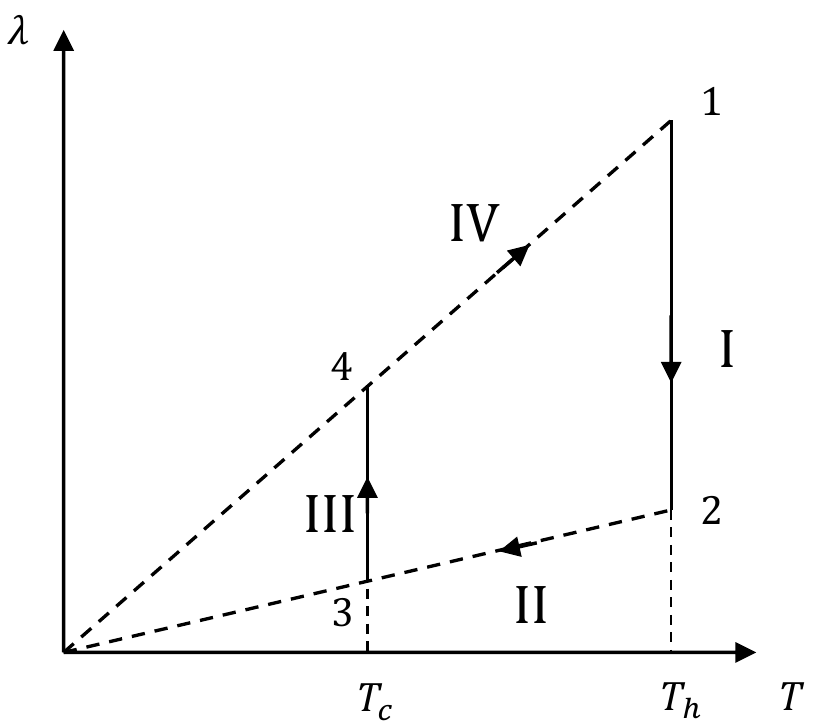}
        \caption{\label{fig:cycle} Carnot-like thermodynamic cycle.}
    \end{figure}

    We construct a microscopic engine with a one-dimensional Brownian particle in a time-dependent harmonic potential. This engine contains two isothermal and two adiabatic branches which are realized by shortcuts to isothermality and shortcuts to adiabaticity, respectively. The original potential is $U_o=\lambda^2(t)x^2/2$. $U_a$ and $H_c$ are respectively applied to the particle during the isothermal and the adiabatic branches. Fig.~\ref{fig:cycle} is a schematic diagram of the thermodynamic cycle. Stage I ranging from time $t_1$ to $t_2$ represents the isothermal expansion branch where the particle is coupled to the hot bath with temperature $T_h$ and $\lambda$ decreases with time. Stage II from time $t_2$ to $t_3$ represents the adiabatic expansion branch where the particle is decoupled from the heat bath. Stage III from time $t_3$ to $t_4$ represents the isothermal compression branch where the particle is coupled to the cold bath with temperature $T_c(<T_h$) and $\lambda$ increases with time. Stage IV represents the adiabatic compression branch after which the particle is again coupled to the hot bath. 

    Stage I is realized by shortcuts to isothermality. According to Eq.~(\ref{Wiso}), the input work during this stage is expressed as
    \begin{equation}
        W_{\mathrm{I}}=T_h\ln\frac{\lambda_2}{\lambda_1}+T_h\frac{C_h[\Lambda_h(\tilde{t})]}{t_2-t_1}
    \end{equation}
    where
    \begin{equation}
        \label{Ch-func}
        C_h[\Lambda_h(\tilde{t})]\equiv\int_0^1\dif{\tilde{t}}\left(\frac{1}{\gamma_h}+\frac{\gamma_h}{\Lambda_h^2}\right)\frac{1}{\Lambda_h^2}\left(\frac{\dif{\Lambda_h}}{\dif{\tilde{t}}}\right)^2,
    \end{equation}
    and 
    $
        \Lambda_h(\tilde{t})\equiv\lambda((t_2-t_1)\tilde{t}+t_1)
    $.
    $\gamma_h$ is the damping coefficient of the particle in hot bath. $\lambda_1$ and $\lambda_2$ are respectively the value of $\lambda$ at the initial and the finial time of stage I. Since the initial and final states of this stage are equilibrium states with same temperature $T_h$, we obtain that the energy difference between them vanishes.
    Based on the conservation of energy, the heat absorbed from the hot bath is
    \begin{equation}
        \label{QI}
            Q_{\mathrm{I}}=-W_{\mathrm{I}}=T_h\ln\frac{\lambda_1}{\lambda_2}-T_h\frac{C_h[\Lambda_h(\tilde{t})]}{t_2-t_1}.
    \end{equation}

    Similarly, the heat exchange between the particle and the cold bath during stage III may be expressed as
    \begin{equation}
        \label{QIII}
        Q_{\mathrm{III}}=T_c\ln\frac{\lambda_3}{\lambda_4}-T_c\frac{C_c[\Lambda_c(\tilde{t})]}{t_4-t_3},
    \end{equation}
    where 
    \begin{equation}
        \label{Cc-func}
        C_c[\Lambda_c(\tilde{t})]\equiv\int_0^1\dif{\tilde{t}}\left(\frac{1}{\gamma_c}+\frac{\gamma_c}{\Lambda_c^2}\right)\frac{1}{\Lambda_c^2}\left(\frac{\dif{\Lambda_c}}{\dif{\tilde{t}}}\right)^2,
    \end{equation}
    and
    $\Lambda_c(\tilde{t})\equiv\lambda((t_4-t_3)\tilde{t}+t_3)$. $\gamma_c$ is the damping coefficient of the particle in cold bath. $\lambda_3$ and $\lambda_4$ are respectively the value of $\lambda$ at initial and finial time of stage III. 
    
    It should be emphasized that $\ln(\lambda_1/\lambda_2)$ in Eq.~(\ref{QI}) and $\ln(\lambda_3/\lambda_4)$ in Eq.~(\ref{QIII}) are exactly the variations of entropy during the isothermal expansion and compression branches, respectively. Considering that stages II and IV are realized by shortcuts to adiabaticity, we have $\lambda_2/T_h=\lambda_3/T_c$ and $\lambda_4/T_c=\lambda_1/T_h$ according to Eq.~(\ref{adipro}). Hence, 
    \begin{equation}
        \frac{\lambda_1}{\lambda_2}=\frac{\lambda_4}{\lambda_3},
    \end{equation}
    which implies that the variations of entropy in the two isothermal branches are opposite numbers. Then Eq.~(\ref{QI}) and Eq.~(\ref{QIII}) can be further expressed as
    \begin{equation}
        Q_{\mathrm{I}}=T_h\left(\Delta S-\frac{C_h}{\tau_h}\right),
    \end{equation}
    and
    \begin{equation}
        Q_{\mathrm{III}}=-T_c\left(\Delta S+\frac{C_c}{\tau_c}\right),
    \end{equation}
    where $\Delta S=\ln(\lambda_1/\lambda_2)=-\ln(\lambda_3/\lambda_4)$ represents the variation of entropy during the isothermal expansion branch. $C_h$ and $C_c$ are respectively the values of the functionals in Eq.~(\ref{Ch-func}) and~(\ref{Cc-func}). We have set $\tau_h\equiv t_2-t_1$ and $\tau_c\equiv t_4-t_3$ for convenience. Noticing that the dissipative terms (i.e. the second term in $Q_{\mathrm{I}}$ and $Q_{\mathrm{III}}$) in the expressions of heat exchanges are inversely proportional to the operation time, we conclude that the microscopic engine based on shortcuts to isothermality and shortcuts to adiabaticity is a realization of microscopic low-dissipation heat engines~\cite{espositoEfficiencyMaximumPower2010}. $C_h$ and $C_c$ respectively correspond to the dissipation coefficients during the isothermal expansion and compression branches in Ref~\cite{espositoEfficiencyMaximumPower2010}. 

    Since the particle will return to the initial state after finishes a thermodynamic cycle, the variation of energy in the cycle vanishes. And there is no heat exchange in the adiabatic branches. Hence, the total output work in each cycle is
    \begin{equation}
        W_{\mathrm{out}}=Q_{\mathrm{I}}+Q_{\mathrm{III}}
    \end{equation}
    with the consideration of energy conservation.

    \section{\label{Sec:IV}EFFICIENCY AT MAXIMUM POWER}
    Considering that the adiabatic branches could be accomplished in rather shorter time than the isothermal branches, the total time for completing the cycle can be approximated by $\tau_h+\tau_c$. Thus, the power output of this microscopic engine is
    \begin{equation}
        \label{power}
            P=\frac{W_{\mathrm{out}}}{\tau_h+\tau_c}=\frac{T_h\left(\Delta S-\frac{C_h}{\tau_h}\right)-T_c\left(\Delta S+\frac{C_c}{\tau_c}\right)}{\tau_h+\tau_c}.
    \end{equation}
    The power output~(\ref{power}) can be optimized with respect to both the protocol of external parameter $\lambda(t)$ and the time $\tau_h$ and $\tau_c$. 
    
\begin{figure}[!htpb]
        \includegraphics[width=0.48\textwidth]{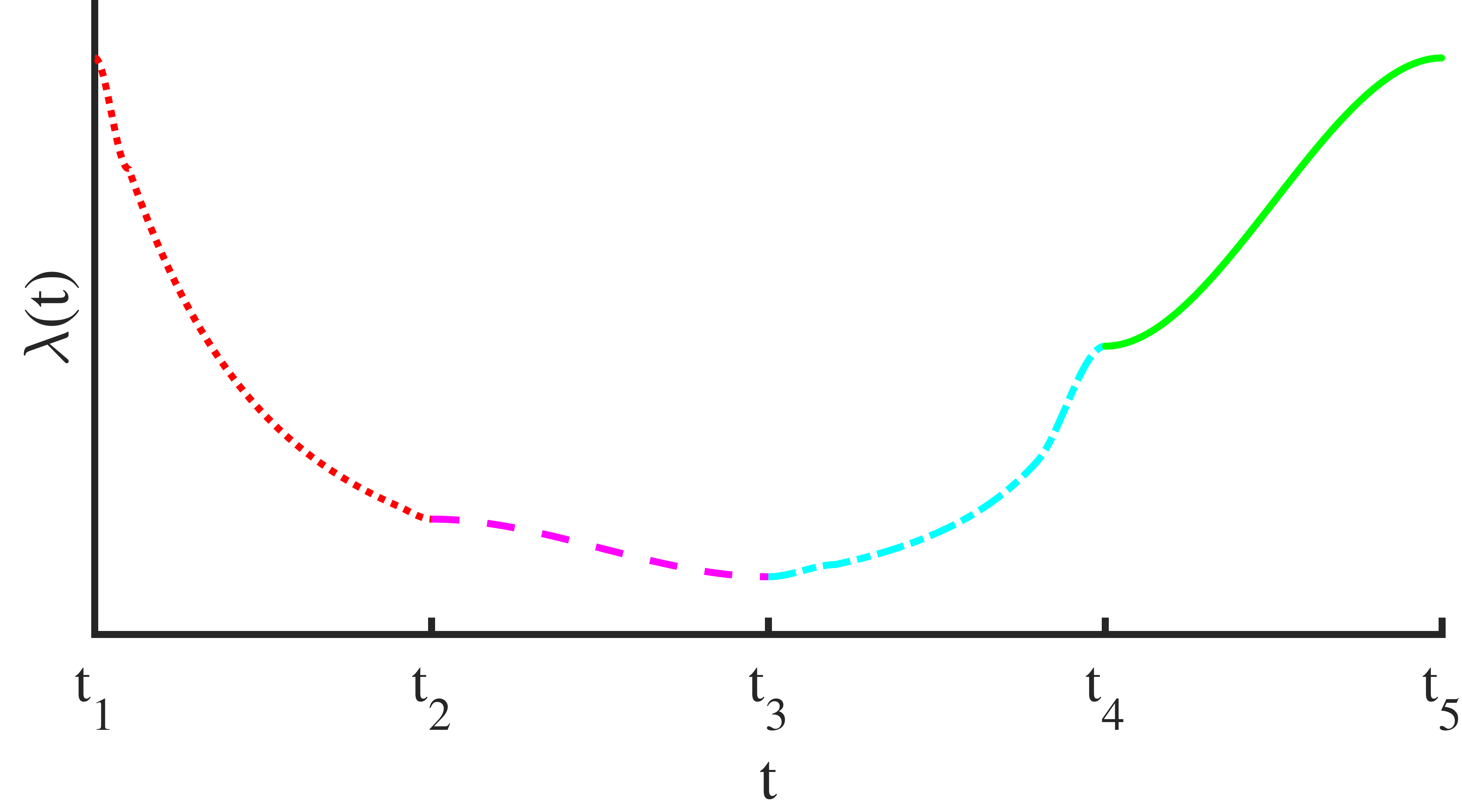}
        \caption{\label{fig:protocol} (Color online) Schematic diagram of the protocol of external parameter. The dotted, dashed, dash-dotted and solid lines respectively represent the isothermal expansion, adiabatic expansion, isothermal compression and adiabatic compression branches. } 
    \end{figure}

    Firstly, to obtain the maximum work output, we need to minimize the functionals $C_{h}[\Lambda_{h}(\tilde{t})]$ and $C_{c}[\Lambda_{c}(\tilde{t})]$. We notice that the integrands in Eq.~(\ref{Ch-func}) and Eq.~(\ref{Cc-func}) do not explicitly contain the argument of time. Hence, there are two conserved quantities:
    \begin{align}
        \Sigma_{h}&\equiv\left(\frac{1}{\gamma_h}+\frac{\gamma_h}{\Lambda_h^2}\right)\frac{1}{\Lambda_h^2}\left(\frac{\dif{\Lambda_h}}{\dif{\tilde{t}}}\right)^2,\label{Sigma-h}\\
        \Sigma_{c}&\equiv\left(\frac{1}{\gamma_c}+\frac{\gamma_c}{\Lambda_c^2}\right)\frac{1}{\Lambda_c^2}\left(\frac{\dif{\Lambda_c}}{\dif{\tilde{t}}}\right)^2.\label{Sigma-c}
    \end{align}
    Further, the minimum values of functionals in Eq.~(\ref{Ch-func}) and Eq.~(\ref{Cc-func}) equal to $\Sigma_h$ and $\Sigma_c$, respectively. We can determine the optimal protocols $\Lambda^*_h$ and $\Lambda^*_c$ in isothermal branches by solving Eq.~(\ref{Sigma-h}) and Eq.~(\ref{Sigma-c}). The detailed calculations can be found in Appendix~\ref{appendix:A}. In adiabatic branches, there is arbitrariness for selecting the protocols. Following Ref.~\cite{tuStochasticHeatEngine2014a}, here we choose $\lambda(t)=\lambda_i+(\lambda_f-\lambda_i)\Phi\left[(t-t_i)/(t_f-t_i))\right]$, where $\Phi(t)$ is defined as $\Phi\equiv3t^2-2t^3$. A schematic diagram of the protocol in a cycle is shown in Fig.~\ref{fig:protocol}. 

    Secondly, by solving $\partial P/\partial \tau_h=0$ and $\partial P/\partial \tau_c=0$, we obtain the optimum values of $\tau_h$ and $\tau_c$:
    \begin{equation}
        \label{optim-th}
        \left\{
            \begin{aligned}
                \tau_h^*&=\frac{2(\sqrt{\Sigma_h \Sigma_c T_hT_c}+\Sigma_hT_h)}{(T_h-T_c)\Delta S},\\
                \tau_c^*&=\frac{2(\sqrt{\Sigma_h \Sigma_c T_hT_c}+\Sigma_cT_c)}{(T_h-T_c)\Delta S},\\
            \end{aligned}
        \right.
    \end{equation}
    where we have replaced $C_h$ and $C_c$ with their optimal values $\Sigma_h$ and $\Sigma_c$, respectively.
    Substituting Eq.~(\ref{optim-th}) into the expression of power, then we obtain the maximum power
    \begin{equation}
        \label{powermax}
        P_{\mathrm{max}}=\frac{T_h\Delta S^2}{\Sigma_h}\frac{\eta_{\C}^2}{4(\sqrt{(1-\eta_{\C})\chi}+1)^2},
    \end{equation}
    where $\chi\equiv\Sigma_c/\Sigma_h$. 
    Based on the definition of efficiency $\eta=W_{\mathrm{out}}/Q_{\mathrm{I}}$, we derive the efficiency at maximum power:
    \begin{equation}
        \label{EMP}
        \eta_{P_{\mathrm{max}}}=\frac{\eta_\mathrm{C}}{2-\frac{\eta_{\mathrm{C}}}{\sqrt{\chi(1-\eta_{\mathrm{C}})}+1}}.
    \end{equation}
    It is not hard to verify that $\eta_{\C}/2\le\eta_{P_{\mathrm{max}}}\le\eta_{\C}/(2-\eta_{\C})$. This constraint is consistent with the conclusion in Ref.~\cite{espositoEfficiencyMaximumPower2010}. Different from the model in Ref.~\cite{espositoEfficiencyMaximumPower2010}, here we can obtain the exact expressions of $\Sigma_h$ and $\Sigma_c$ according to differential equations~(\ref{Sigma-h}) and~(\ref{Sigma-c}):

    \begin{align}
        \Sigma_h&=\frac{1}{\gamma_h}\left[\psi\left(\frac{\lambda_2}{\gamma_h}\right)-\psi\left(\frac{\lambda_1}{\gamma_h}\right)\right]^2,\label{Sigma-h-sol}\\
        \Sigma_c&=\frac{1}{\gamma_c}\left[\psi\left(\frac{\lambda_4}{\gamma_c}\right)-\psi\left(\frac{\lambda_3}{\gamma_c}\right)\right]^2,\label{Sigma-c-sol}
    \end{align}
    with $\psi(x)\equiv\sinh^{-1}(x)-\cosh[\sinh^{-1}(1/x)]$. The detailed derivation can be found in Appendix~\ref{appendix:A}. Then we have 
    \begin{equation}
        \label{chi}
        \chi=\frac{\Sigma_c}{\Sigma_h}=\xi\frac{\Psi((1-\eta_{\C})\xi\alpha)}{\Psi(\alpha)},
    \end{equation}
    where $\xi\equiv\gamma_h/\gamma_c$, $\alpha\equiv\lambda_1/\gamma_h$ and $\Psi(x)\equiv\left[\psi(\zeta x)-\psi(x)\right]^2$ with $\zeta\equiv\lambda_2/\lambda_1$. 
    
\begin{figure}[!htpb]
        \includegraphics[width=0.48\textwidth]{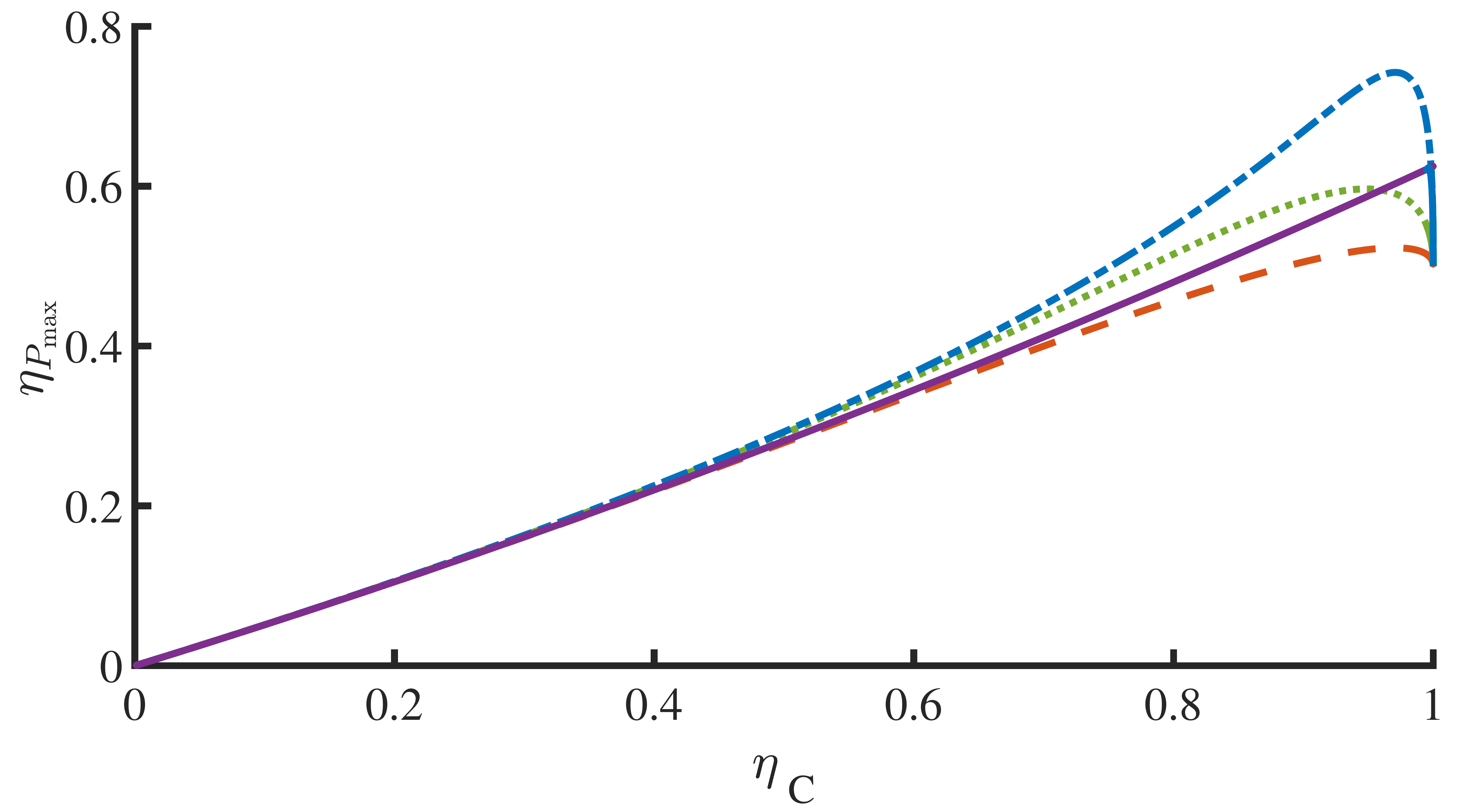}
        \caption{\label{fig:EMPline} (Color online) Efficiency at maximum power with $\xi=1$ and $\zeta=0.3$. The dashed, dotted and dash-dotted lines respectively correspond to Eq.~(\ref{EMP}) with $\alpha=\pi/10$, $\alpha=2\pi$ and $\alpha=10\pi$. The solid line corresponds to Eq.~(\ref{emp-quadratic}).} 
    \end{figure}

    It is obvious that the efficiency at maximum power depends on the parameters $\alpha$, $\zeta$ and $\xi$ as well as the Carnot efficiency $\eta_{\C}$. For symmetric damping situation where $\xi=1$, Eq.~(\ref{chi}) becomes $\chi\approx 1-[\alpha\Psi^{\prime}(\alpha)/\Psi(\alpha)]\eta_{\C}$ for small $\eta_{\C}$. By substituting $\chi$ into Eq.~(\ref{EMP}), we obtain the efficiency at maximum power up to the quadratic order of $\eta_{\C}$
    \begin{equation}
        \label{emp-quadratic}
        \eta_{P_{\mathrm{max}}}\approx\frac{\eta_{\mathrm{C}}}{2}+\frac{\eta_{\mathrm{C}}^2}{8},
    \end{equation}
    which is consistent with the universal law of efficiency at maximum power~\cite{espositoUniversalityEfficiencyMaximum2009,shengConstitutiveRelationNonlinear2015,tuAbstractModelsHeat2021}. In fact, the damping coefficients are usually dependent of temperature, which causes $\xi$ deviates from $1$. However, we find that $\xi=1+O(\eta_{\C})$ for most of solvents~\cite{viswanathViscosityLiquidsTheory2006,haj-kacemContributionModelingViscosity2014a,messaadiNewEquationRelating2015}. In this case, the universal law~(\ref{emp-quadratic}) still holds. In Fig.~\ref{fig:EMPline}, we compare the behavior of Eq.~(\ref{EMP}) with that of Eq.~(\ref{emp-quadratic}) for different $\alpha$ when $\xi=1$ and $\zeta=0.3$. The dashed, dotted and dash-dotted lines respectively correspond to Eq.~(\ref{EMP}) with $\alpha=\pi/10$, $\alpha=2\pi$ and $\alpha=10\pi$. The solid line corresponds to the universal law~(\ref{emp-quadratic}). We observe that these curves overlap at small $\eta_{\C}$. 

    Moreover, we notice that the efficiency at maximum power given by Eq.~(\ref{EMP}) tends to $1/2$ when $\eta_{\C}$ tends to $1$. This surprising result is different from the performances of endoreversible Carnot-like engine~\cite{curzonEfficiencyCarnotEngine1975,chenEffectHeatTransfer1989}, stochastic engine~\cite{t.schmiedlEfficiencyMaximumPower2007}, Feynman's ratchet~\cite{tuEfficiencyMaximumPower2008}, quantum dot engine~\cite{espositoThermoelectricEfficiencyMaximum2009} etc. To understand this result, we  calculate the Laurent series of $\chi$ about $\eta_{\C}=1$ when $\xi =1$:
    \begin{equation}
        \label{chi-series}
        \chi=\frac{(1-\zeta)^2}{\zeta^2\alpha^2\Psi(\alpha)\left(1-\eta_{\C}\right)^2}+O((1-\eta_{\C})^0).
    \end{equation}
    Substituting this equation into Eq.~(\ref{EMP}), then we obtain
    \begin{equation}
        \eta_{P_{\mathrm{max}}}=\frac{1}{2}+\frac{\zeta\alpha\sqrt{\Psi(\alpha)}}{4(1-\zeta)}\sqrt{1-\eta_{\C}}+O((1-\eta_{\C})^1).
    \end{equation}
    The above equation describes the behavior of the three curves corresponding to Eq.~(\ref{EMP}) nearby $\eta_{\C}=1$ in Fig.~\ref{fig:EMPline}.
    
    \section{\label{Sec:V}POWER AT ANY GIVEN EFFICIENCY}
    Since the efficiency and power could not be simultaneously optimized, we need to investigate the trade-off relation between these two quantities. For asymmetrical damping coefficients and anharmonic potentials, the microscopic model in the present work becomes a general microscopic low-dissipation engine. In this section, we start with general low-dissipation engines and find the maximum power at given efficiency.

    To simplify the calculations, we define the following quantities: $\alpha_{h}=\Sigma_{h}/\Delta S$, $\alpha_{c}=\Sigma_{c}/\Delta S$, $L_h=\alpha_h /\tau_h$, $L_c=\alpha_c /\tau_c$, $\tilde{\tau}=\tau_h/\tau_c$. The expressions of heat exchanges $Q_{\mathrm{I}}$ and $Q_{\mathrm{III}}$ with the new variables are respectively
    \begin{equation}
        Q_{\mathrm{I}}=\Delta S T_h (1-L_h),
    \end{equation}
    and
    \begin{equation}
        Q_{\mathrm{III}}=-\Delta S T_c(1+L_c).
    \end{equation}
    Hence, the efficiency of the engine is
    \begin{equation}
        \label{eta-new-variable}
        \eta=\frac{Q_{\mathrm{I}}+Q_{\mathrm{III}}}{Q_{\mathrm{I}}}=1-(1-\eta_\mathrm{C})\frac{1+\chi\tilde{\tau}L_h}{1-L_h},
    \end{equation}
    where we have used $L_c/L_h=\chi\tilde{\tau}$. From Eq.~(\ref{eta-new-variable}), we obtain
    \begin{equation}
        \tilde{\tau}=b\left(\frac{\delta}{L_h}-1\right),
    \end{equation}
    where
    \begin{equation}
        \label{b-chi}
        b=\frac{1-\eta}{(1-\eta_\mathrm{C})\chi},
    \end{equation}
    and 
    \begin{equation}
        \label{delta}
        \delta=\frac{\eta_{\C}-\eta}{1-\eta}\le\eta_{\C}.
    \end{equation}
    Since $b\ge 0$ and $\tilde{\tau}\ge0$, we obtain $L_h\le\delta\le \eta_{\C}$. The power of the engine is
    \begin{equation}
        \label{P-eta}
        P=\frac{\eta Q_{\mathrm{I}}}{\tau_h+\tau_c}=\eta\Delta S\frac{T_h}{\alpha_h}\frac{L_h(1-L_h)(\delta-L_h)}{\delta-L_h+L_h/b}.
    \end{equation}
    In the following, we consider the dimensionless power
    \begin{equation}
        \label{Ptilde-eta}
        \tilde{P}=\frac{\alpha_h P}{T_h \Delta S}=\eta\frac{L_h(1-L_h)(\delta-L_h)}{\delta-L_h+L_h/b}.
    \end{equation} 
    To obtain the maximum value of $\tilde{P}$ for given $\eta$, we need to solve the equation $\partial \tilde{P}/\partial L_h=0$ for $L_h$. However, this is a cubic equation and the solutions are so cumbersome that we can not illustrate their analytical behavior. Hence, we consider three limiting cases as follows.

\begin{figure}[!htbp]
        \includegraphics[width=0.48\textwidth]{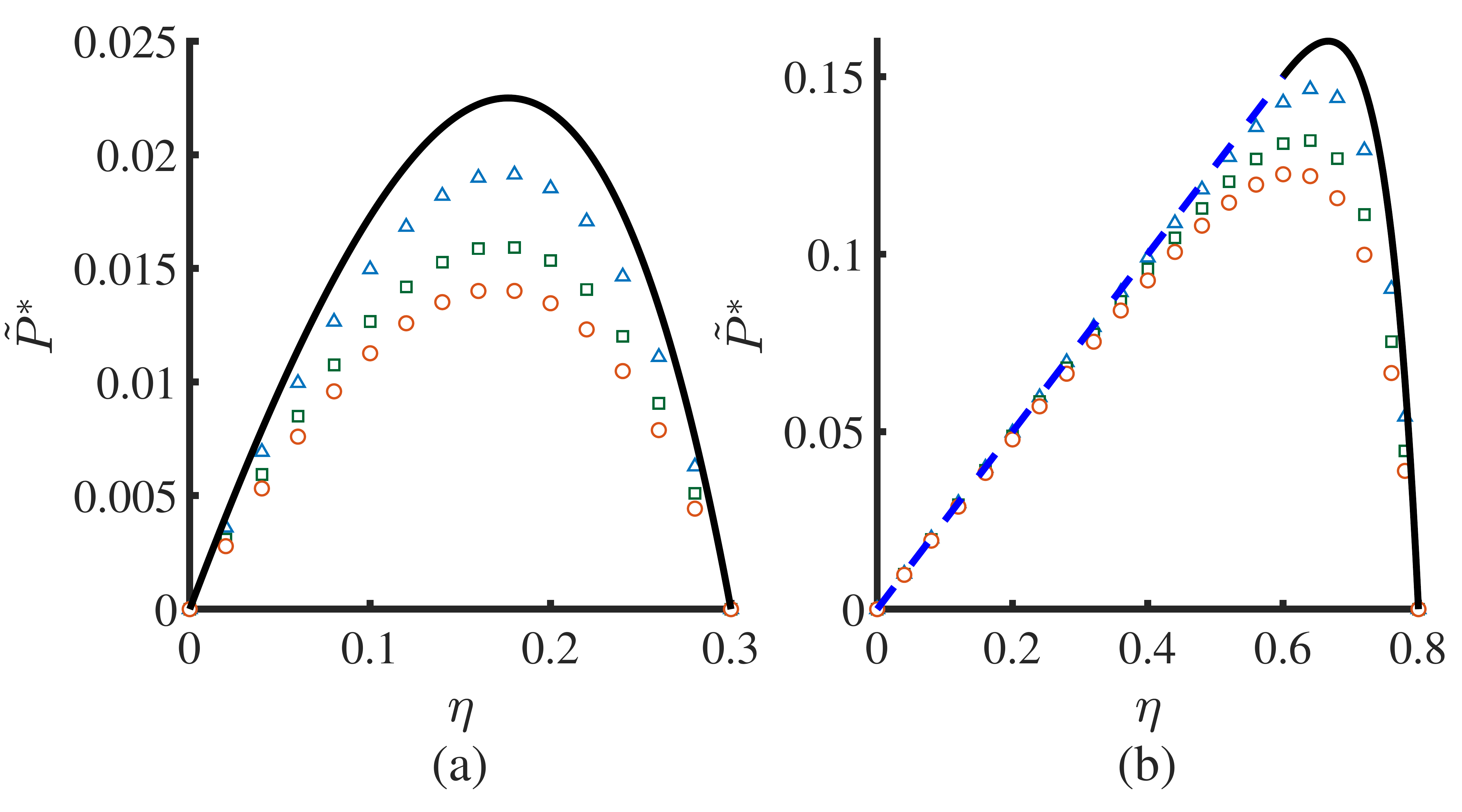}
        \caption{\label{fig:Peta-zero} (color online) The upper bound of power at given efficiency for small $\chi$. The dashed and solid lines correspond to Eq.~(\ref{Peta0m-1}) and Eq.~(\ref{Peta0m-2}), respectively. The triangles, squares and circles respectively represent the numerical upper bounds of power for $\chi=0.01$, $\chi=0.05$ and $\chi=0.1$. (a) $\eta_{\C}=0.3$; (b) $\eta_{\C}=0.8$.}
    \end{figure}

    Firstly, $\chi\equiv\Sigma_c/\Sigma_h\to 0$ which means the dissipation is dominated by the isothermal expansion branch. In this case, Eq.~(\ref{b-chi}) implies $b\to\infty$. Hence, Eq.~(\ref{Ptilde-eta}) degenerates into
    \begin{equation}
        \label{Peta0}
        \tilde{P}_{0}=\eta L_h(1-L_h).
    \end{equation}
    For given efficiency, this is a parabolic function of $L_h$ with the extreme point at $L_h=1/2$. Since $L_h\le\delta\le \eta_{\C}$, the accessible maximum value of $\tilde{P}_{0}$ is dependent of $\delta$. 
    If $\delta\ge 1/2$, which means $\eta\le 2\eta_{\C}-1$, $\tilde{P}_{0}$ reaches its maximum value at $L_h=1/2$ and the value is
    \begin{equation}
        \label{Peta0m-1}
        \tilde{P}_{0}^*=\frac{\eta}{4}.
    \end{equation}
    If $\delta< 1/2$, which means $\eta> 2\eta_{\C}-1$, $\tilde{P}_{0}$ reaches its maximum value at $L_h=\delta$ and the value is
    \begin{equation}
        \label{Peta0m-2}
        \tilde{P}_{0}^*=\frac{\eta(1-\eta_\mathrm{C})(\eta_\mathrm{C}-\eta)}{(1-\eta)^2}.
    \end{equation}
    The global maximum value of $\tilde{P}_{0}^*$ is $\eta_\C^2/4$ at $\eta=\eta_\C/(2-\eta_\C)$. We emphasize that the value of $\eta_{\C}$ determines whether the maximum power at given efficiency can be expressed as Eq.~(\ref{Peta0m-1}). If $\eta_{\C}\le 1/2$, Eq.~(\ref{delta}) implies $\delta\le 1/2$ for $\eta\in [0,\eta_{\C}]$. Then the maximum power at given efficiency is only expressed as Eq.~(\ref{Peta0m-2}). If $\eta_{\C}>1/2$, Eq.~(\ref{delta}) implies that $\delta$ can be either larger or smaller than $1/2$. Then the maximum power at given efficiency is piecewisely expressed as Eq.~(\ref{Peta0m-1}) and Eq.~(\ref{Peta0m-2}). Fig.~\ref{fig:Peta-zero} shows the upper bound of power at given efficiency for small $\chi$ when $\eta_{\C}=0.3$ (Fig.~\ref{fig:Peta-zero}(a)) and $\eta_{\C}=0.8$ (Fig.~\ref{fig:Peta-zero}(b)). The dashed and solid lines correspond to the analytical results~(\ref{Peta0m-1}) and~(\ref{Peta0m-2}), respectively. The triangles, squares and circles respectively represent the upper bounds of power for $\chi=0.01$, $\chi=0.05$ and $\chi=0.1$ obtained with numerical method (see Appendix~\ref{appendix:B}). From Fig.~\ref{fig:Peta-zero}, we observe that the numerical results approach the analytical result as $\chi$ decreases.

\begin{figure}[!htbp]
        \includegraphics[width=0.48\textwidth]{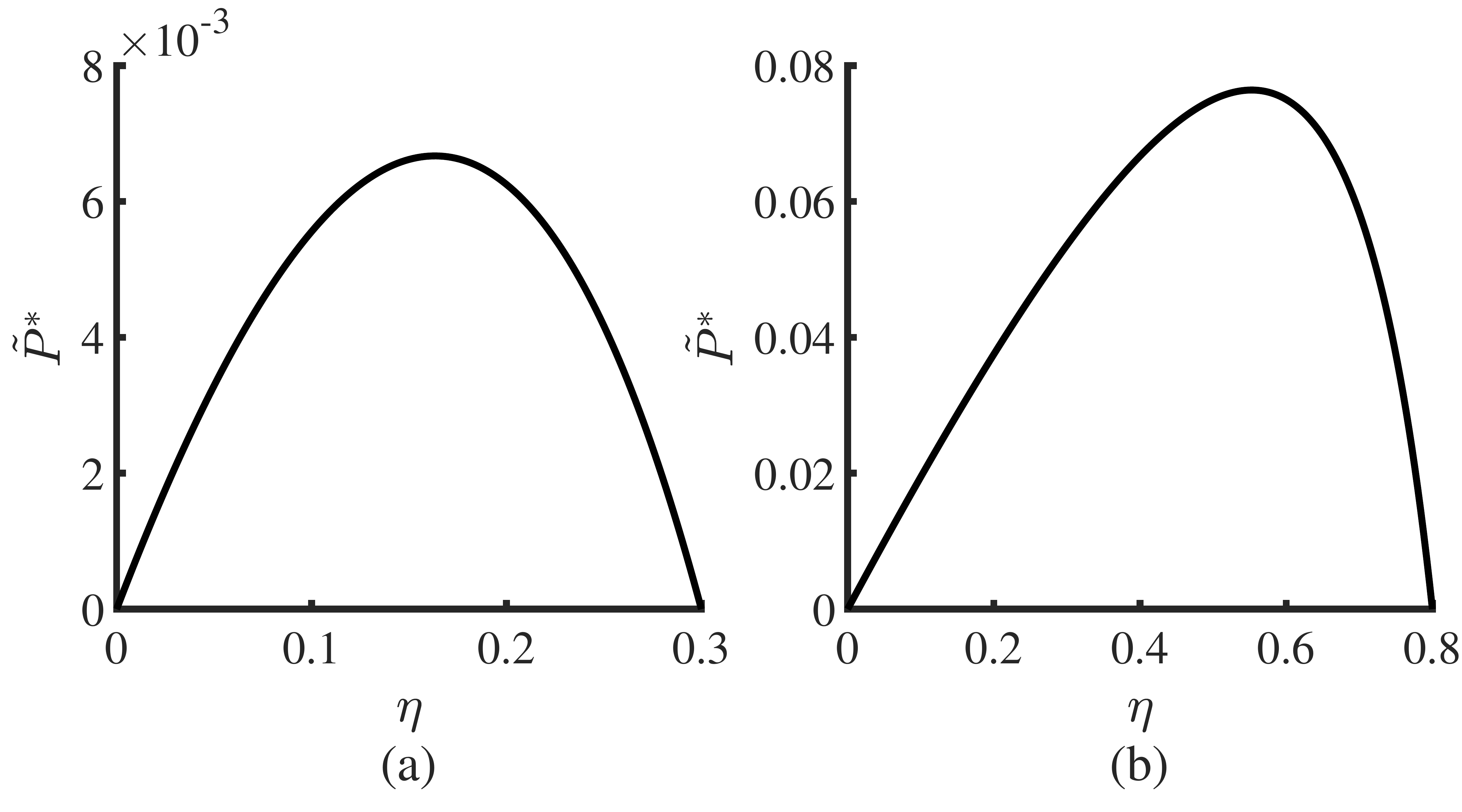}
        \caption{\label{fig:Peta-one} The upper bounds of power at given efficiency for $\chi=1$ according to Eq.~(\ref{Peta1m}). (a) $\eta_{\C}=0.3$; (b) $\eta_{\C}=0.8$. }
    \end{figure}

    Secondly, $\chi= 1$ which implies that the dissipation coefficients are symmetric in the two isothermal branches. In this case, Eq.~(\ref{b-chi}) implies $b=(1-\eta)/(1-\eta_\mathrm{C})$. Hence, Eq.~(\ref{Ptilde-eta}) degenerates into
    \begin{equation}
        \label{Peta1}
            \tilde{P}_{1}=\eta L_h(\delta-L_h)/\delta.
    \end{equation}
    Obviously, for given efficiency, $\tilde{P}_{1}$ reaches its maximum value at $L_h=\delta/2$. The maximum value is
    \begin{equation}
        \label{Peta1m}
        \tilde{P}_{1}^*=\frac{\eta(\eta_\mathrm{C}-\eta)}{4(1-\eta)}.
    \end{equation} 
    This expression is the same as the result of endoreversible engine obtained by Chen and Yan~\cite{chenEffectHeatTransfer1989}. The global maximum value of $\tilde{P}_{1}^*$ is $\left(1-\sqrt{1-\eta_\C}\right)^2/4$ at $\eta=1-\sqrt{1-\eta_\C}$ which is exactly the efficiency at maximum power obtained by Curzon and Ahlborn~\cite{curzonEfficiencyCarnotEngine1975}. Fig.~\ref{fig:Peta-one} shows the upper bounds~(\ref{Peta1m}) of power at given efficiency for $\chi=1$. 
    
\begin{figure}[!htbp]
        \includegraphics[width=0.48\textwidth]{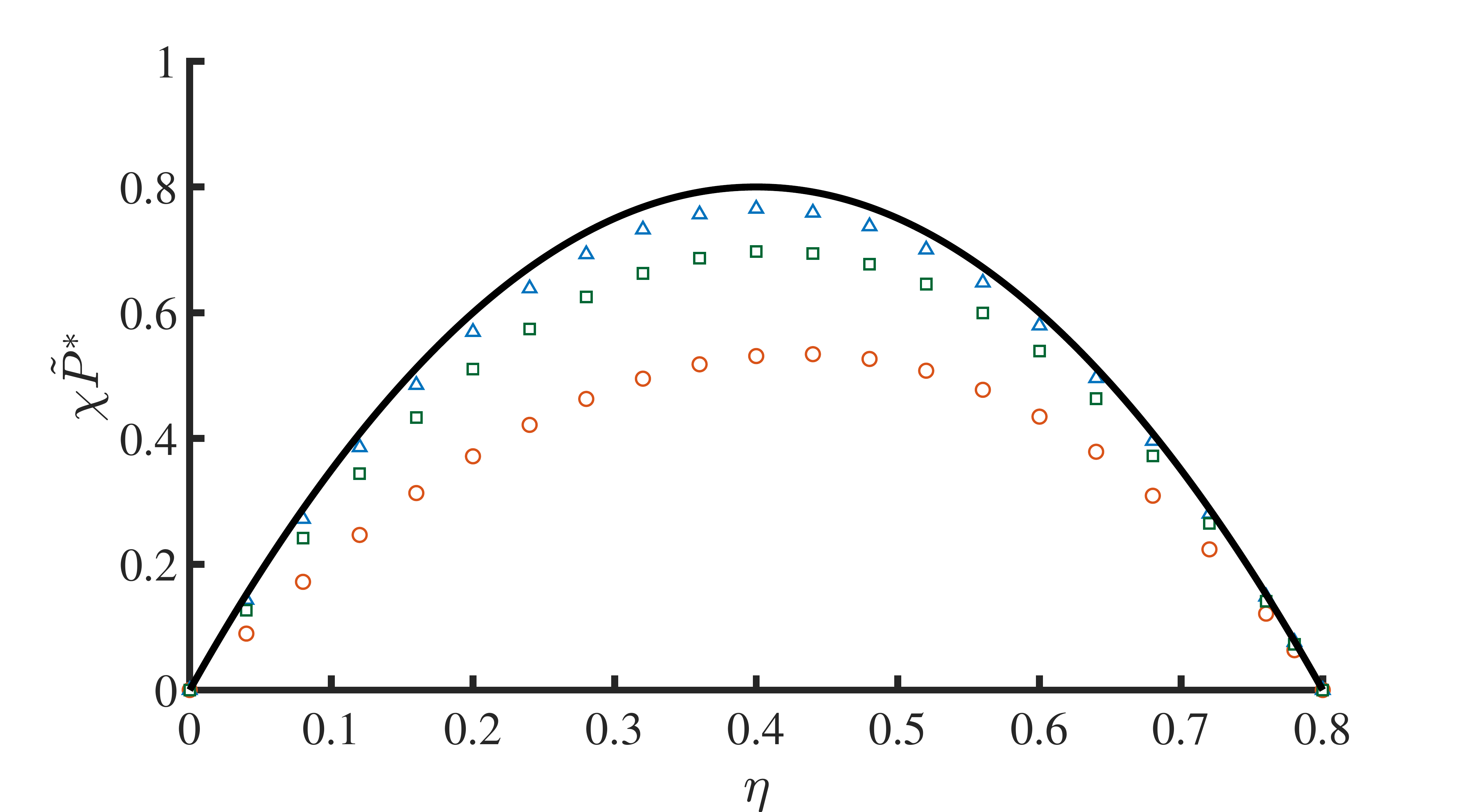}
        \caption{\label{fig:Peta-inf} (color online) The product of $\chi$ and the maximum power at given efficiency for large $\chi$. The solid line corresponds to the analytical result based on Eq.~(\ref{Petainftym}). The circles, squares and triangles respectively correspond to the numerical results for $\chi=10^2$, $\chi=10^3$ and $\chi=10^4$. $\eta_{\C}=0.8$.}
    \end{figure}

    Thirdly, $\chi\to \infty$ which means the dissipation is dominated by the isothermal compression branch. In this case, Eq.~(\ref{b-chi}) implies $b\to 0$. The leading term of Eq.~(\ref{Ptilde-eta}) is
    \begin{equation}
        \label{Petainfty}
        \tilde{P}_{\infty}=\frac{1}{\chi_{\infty}}\frac{\eta(1-\eta)}{1-\eta_\mathrm{C}}(1-L_h)(\delta-L_h),
    \end{equation}
    where the subscript `$\infty$' of $\chi_{\infty}$ indicates that $\chi$ is sufficiently large. For given efficiency, the maximum value of $\tilde{P}_{\infty}$ is attained when $L_h=0$ and the maximum value is
    \begin{equation}
        \label{Petainftym}
        \tilde{P}_{\infty}^*=\frac{1}{\chi_\infty}\frac{\eta(\eta_\mathrm{C}-\eta)}{1-\eta_\mathrm{C}}.
    \end{equation}
    The global maximum value of $\tilde{P}_{\infty}^*$ is $\eta_\C^2/\left[4\chi_\infty(1-\eta_\C)\right]$ at $\eta=\eta_\C/2$. Fig.~\ref{fig:Peta-inf} shows the behavior of $\chi\tilde{P}^*$ for large $\chi$. The solid line corresponds to the analytical result based on Eq.~(\ref{Petainftym}). The circles, squares and triangles respectively correspond to the numerical results for $\chi=10^2$, $\chi=10^3$ and $\chi=10^4$ (the numerical method can be found in Appendix~\ref{appendix:B}). From Fig.~\ref{fig:Peta-inf}, we observe that the numerical results approach the analytical result as $\chi$ increases.

    \section{\label{Sec:VI}efficiency at any given power}
    In Sec.~\ref{Sec:V}, we have investigated the constraint on power at given efficiency and obtained the maximum-power curves in the power-efficiency diagram. In fact, these curves also provide the constraint on efficiency at given power, which was studied in Refs.~\cite{holubecMaximumEfficiencyLowdissipation2016,maUniversalConstraintEfficiency2018a}. Holubec and Ryabov discussed the upper bound of efficiency at given power in Ref.~\cite{holubecMaximumEfficiencyLowdissipation2016} while Ma {\it et al.} analyzed the upper and lower bounds of efficiency at given power in Ref.~\cite{maUniversalConstraintEfficiency2018a}. In this section, we compare our results in Sec.~\ref{Sec:V} with those in Refs.~\cite{holubecMaximumEfficiencyLowdissipation2016,maUniversalConstraintEfficiency2018a}.

\begin{figure}[!htbp]
        \includegraphics[width=0.45\textwidth]{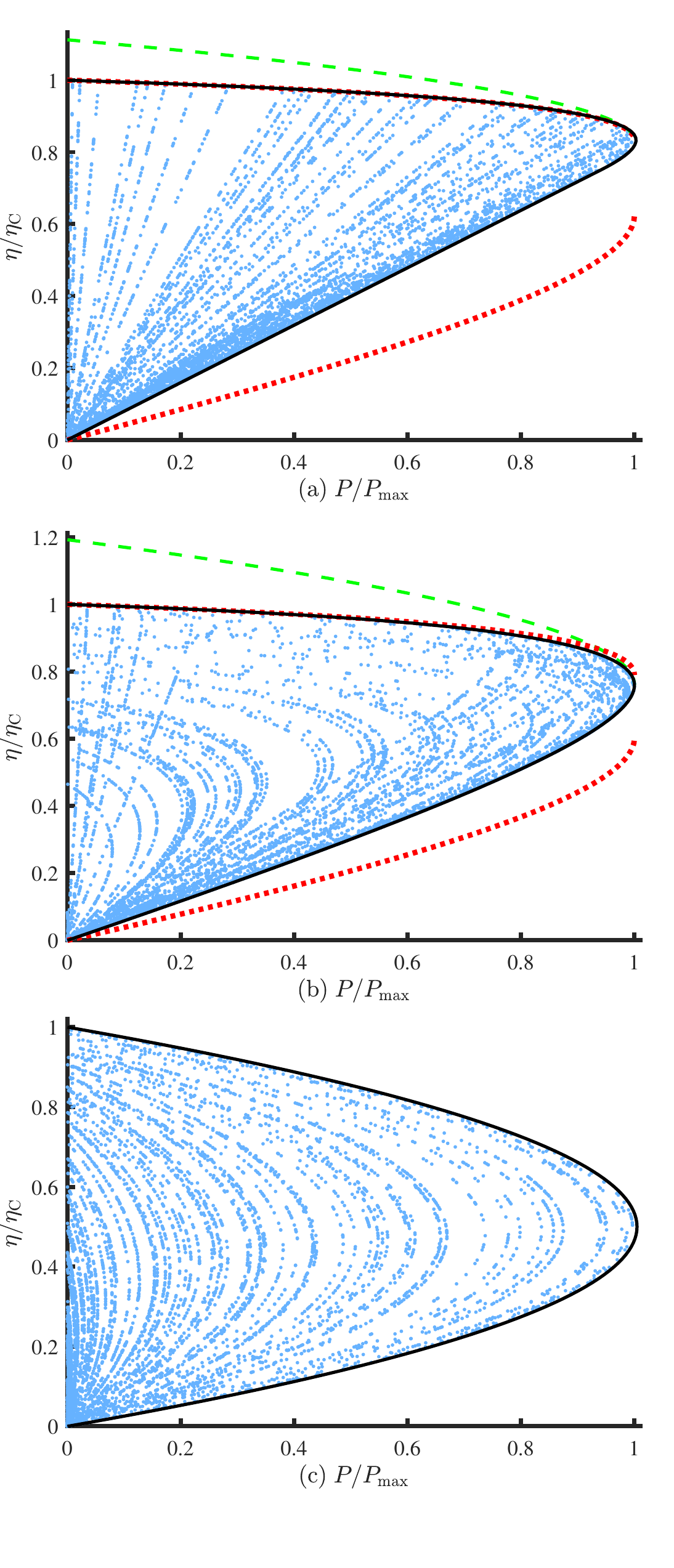}
        \caption{\label{fig:etaP-full} (color online) The power-efficiency diagram of low-dissipation engines. $\eta_{\C}=0.8$ and $P_{\mathrm{max}}$ is given by Eq.~(\ref{powermax}). The scatter points are generated by random values of $L_h$ and $\tilde{\tau}$. The solid lines are based on the results in Sec.~\ref{Sec:V}. The dashed and dotted lines respectively correspond to the results in Ref.~\cite{holubecMaximumEfficiencyLowdissipation2016} and Ref.~\cite{maUniversalConstraintEfficiency2018a}. (a) $\chi=10^{-5}$, (b) $\chi=1$, (c) $\chi=10^6$. }
    \end{figure}

    The power of the engine can be expressed with $\tilde{\tau}$ and $L_h$ as follows:
    \begin{equation}
        \label{power-new-variable}
        \begin{aligned}
            P&=\frac{Q_{\mathrm{I}}+Q_{\mathrm{III}}}{\tau_h+\tau_c}\\
            &=\frac{\Delta S T_h}{\alpha_h}\tilde{\tau}L_h\frac{(1-L_h)-(1-\eta_{\C})(1+\chi\tilde{\tau}L_h)}{1+\tilde{\tau}},
        \end{aligned}
    \end{equation}
    while the efficiency is expressed as Eq.~(\ref{eta-new-variable}). To ensure that the power and the operation time are non-negative, the value of $L_h$ is confined by $0\le L_h\le\eta_{\C}$ and $\tilde{\tau}$ is confined by $0\le\tilde{\tau}\le[(1-L_h)/(1-\eta_{\C})-1]/(\chi L_h)$. Fig.~\ref{fig:etaP-full} shows the power-efficiency diagram of the low-dissipation engines in three limiting cases: $\chi\to 0,1,$ and $\infty$. The scatter points represent the possible values of power and efficiency, which are generated by random values of $L_h$ and $\tilde{\tau}$ according to Eq.~(\ref{eta-new-variable}) and Eq.~(\ref{power-new-variable}). Fig.~\ref{fig:etaP-full}(a) shows the results for $\chi\to 0$. The solid line is depicted according to Eqs.~(\ref{Peta0m-1}) and~(\ref{Peta0m-2}). It is exactly the envelope curve of all scatter points. The dashed line corresponds to the upper bound of efficiency at given power obtained in Ref.~\cite{holubecMaximumEfficiencyLowdissipation2016}. The two dotted lines represent the lower and upper bounds of efficiency at given power obtained in Ref.~\cite{maUniversalConstraintEfficiency2018a}. It is obvious that the upper bound of efficiency in Ref.~\cite{holubecMaximumEfficiencyLowdissipation2016} is a supremum at large power but deviates from the accessible values of efficiency at small power. The upper bound in Ref.~\cite{maUniversalConstraintEfficiency2018a} overlaps with our result while the lower bound is inaccessible. Fig.~\ref{fig:etaP-full}(b) shows the results for $\chi=1$. The solid line is depicted according to Eq.~(\ref{Peta1m}). It is the envelope curve of all scatter points. The upper bound of efficiency in Ref.~\cite{holubecMaximumEfficiencyLowdissipation2016} does well at large power and the upper bound of efficiency in Ref.~\cite{maUniversalConstraintEfficiency2018a} does well for most power values. The lower bound in Ref.~\cite{maUniversalConstraintEfficiency2018a} is still inaccessible. Fig.~\ref{fig:etaP-full}(c) shows the results for $\chi\to\infty$. The solid line as an envelope curve of the scatter points, is depicted according to Eq.~(\ref{Petainftym}) and it overlaps with the constraints in Ref.~\cite{holubecMaximumEfficiencyLowdissipation2016,maUniversalConstraintEfficiency2018a}. 

    Based on the above discussion, we conclude that the results in Sec.~\ref{Sec:V} provide more exact constraints for efficiency and power than the bounds in Ref.~\cite{holubecMaximumEfficiencyLowdissipation2016,maUniversalConstraintEfficiency2018a} for low-dissipation engines. 

    \section{\label{Sec:VII}CONCLUSION}
    In summary, we have designed a Carnot-like microscopic heat engine with the help of shortcuts to isothermality and shortcuts to adiabaticity. The dissipative work during the isothermal branches is inversely proportional to the operation time, which means we have realized a microscopic low-dissipation engine. Although we have only demonstrated the case of harmonic potential, this realization is not restricted to harmonic potential since the inverse-proportion relation is independent of the form of potentials according to the character of shortcuts to isothermality. We have obtained the analytical efficiency at maximum power of this engine. For symmetric damping coefficients, we have verified that the efficiency at maximum power satisfies the universal law~(\ref{emp-quadratic}) at small $\eta_{\C}$ and tends to a universal value (i.e., $1/2$) when $\eta_{\C}$ approaches one. Our results are different from those obtained by Schmiedl and Seifert~\cite{t.schmiedlEfficiencyMaximumPower2007}. The underlying reason is the consideration of shortcuts to isothermality and shortcuts to adiabaticity in our model. 
    
    For asymmetrical damping coefficients and anharmonic potentials, the microscopic model in the present work becomes a general microscopic low-dissipation engine. We have investigated the maximum power at given efficiency for general low-dissipation engines and derived the analytical results in three limiting cases. When the ratio $\chi$ of the dissipation coefficients during the cold isothermal branch and the hot isothermal branch is one, the result for endoreversible engine is recovered. For $\chi$ approaches zero or infinity, the expression of maximum power at given efficiency becomes fairly concise, which is respectively piecewise curve or parabolic curve. We have compared the constraints for power and efficiency in the present work with those in Refs.~\cite{holubecMaximumEfficiencyLowdissipation2016,maUniversalConstraintEfficiency2018a} and obtained that our constraints are more exact. We notice that in recent work~\cite{chenOptimizingShortcutBrownian2022a}, Chen implemented the strategies of generalized shortcuts to isothermality to design Brownian heat engines and obtained the efficiency at maximum power as well as the maximum power at given efficiency via thermodynamic length. It is worth considering to calculate the thermodynamic length of our model in the future work. In addition, Albay {\it et al}. experimentally realized the shortcuts to isothermality~\cite{albayThermodynamicCostShortcutstoisothermal2019,albayWorkRelationInstantaneousequilibrium2020,albayRealizationFiniterateIsothermal2020}. Hence, it is possible to verify our theoretical results in future experiments.
    
    \begin{acknowledgments}
        The authors are grateful for financial support from the National Natural Science Foundation of China (Grant No. 11975050). We are very grateful for the discussion with Geng Li.
    \end{acknowledgments}

    \appendix

    \section{\label{appendix:A}Detailed discussion about optimal potential protocols}
    Based on the differential equations~(\ref{Sigma-h}) and~(\ref{Sigma-c}), the optimal protocols ($\Lambda_h^*(\tilde{t})$ and $\Lambda_c^*(\tilde{t})$) of external parameters are given by the following implicit expressions:
    \begin{align}
        &\sinh^{-1}\left(\frac{\Lambda_h^*}{\gamma_h}\right)-\cosh\left[\sinh^{-1}\left(\frac{\gamma_h}{\Lambda_h^*}\right)\right]=-\sqrt{\gamma_h\Sigma_h}\tilde{t}+c_1,\label{Lambdah_star}\\
        &\sinh^{-1}\left(\frac{\Lambda_c^*}{\gamma_c}\right)-\cosh\left[\sinh^{-1}\left(\frac{\gamma_c}{\Lambda_c^*}\right)\right]=\sqrt{\gamma_c\Sigma_c}\tilde{t}+c_2,\label{Lambdac_star}
    \end{align}
    where $\Sigma_h$, $c_1$, $\Sigma_c$ and $c_2$ are time-independent constants which can be determined by the boundary conditions $\Lambda_h(0)=\lambda_1$, $\Lambda_h(1)=\lambda_2$, $\Lambda_c(0)=\lambda_3$ and $\Lambda_c(1)=\lambda_4$. The expressions of these constants are:
    \begin{align}
        &c_1=\psi\left(\frac{\lambda_1}{\gamma_h}\right), \Sigma_h=\frac{1}{\gamma_h}\left[\psi\left(\frac{\lambda_2}{\gamma_h}\right)-\psi\left(\frac{\lambda_1}{\gamma_h}\right)\right]^2;\label{const_Lambdah_star}\\
        &c_2=\psi\left(\frac{\lambda_3}{\gamma_c}\right), \Sigma_c=\frac{1}{\gamma_c}\left[\psi\left(\frac{\lambda_4}{\gamma_c}\right)-\psi\left(\frac{\lambda_3}{\gamma_c}\right)\right]^2.\label{const_Lambdac_star}
    \end{align}
    $\psi$ is defined as $\psi(x)\equiv\sinh^{-1}\left(x\right)-\cosh\left[\sinh^{-1}\left(1/x\right)\right]$.
    
    On the other hand, considering that $\dif{\Lambda_h}/\dif{\tilde{t}}=\tau_h\dif{\lambda}/\dif{t}$, the constraint~(\ref{bc}) means that
    \begin{equation}
        \label{bc-dLambda}
        \left.\frac{\dif{\Lambda_h}}{\dif{\tilde{t}}}\right|_{\tilde{t}=0}=\left.\frac{\dif{\Lambda_h}}{\dif{\tilde{t}}}\right|_{\tilde{t}=1}=0.
    \end{equation}
    If we substitute the above conditions into Eq.~(\ref{Sigma-h}), we will obtain $\Sigma_h=0$ (considering that $\Sigma_h$ is a conserved quantity during the isothermal expansion branch). This is incompatible with Eq.~(\ref{const_Lambdah_star}) since $\lambda_1\neq\lambda_2$. There is same contradiction in the isothermal compression branch. To resolve these contradictions, we introduce a small time interval $\varepsilon$ and turn the protocols into piecewise functions over $\tilde{t}\in[0,1]$. Take the isothermal expansion branch as an example.\\
     For $\tilde{t}\in[0,\varepsilon]$, we suppose that
    \begin{equation}
        \label{Lambda*l}
        \Lambda_{h}^{*l}(\tilde{t})=\Lambda_h(0)+\left[a_h^l-\Lambda_h(0)\right]\phi_h^l\left(\frac{\tilde{t}}{\varepsilon}\right),
    \end{equation}
    where $\phi_h^l(t)$ is defined as $\phi_h^l(t)=b_h^lt^2-(b_h^l-1)t^3$, $a_h^l$ and $b_h^l$ are undetermined coefficients. It is easy to verify that $\Lambda_h^{*l}(\tilde{t})$ satisfies the boundary conditions of $\Lambda_h$ and $\dif{\Lambda_h}/\dif{\tilde{t}}$ at $\tilde{t}=0$.\\
    For $\tilde{t}\in[\varepsilon,1-\varepsilon]$, 
    \begin{equation}
        \label{Lambda*m}
        \Lambda_{h}^{*m}(\tilde{t})=\Lambda_{h}^*.
    \end{equation}
    \\
    For $\tilde{t}\in[1-\varepsilon,1]$, we suppose that
    \begin{equation}
        \label{Lambda*r}
        \Lambda_{h}^{*r}(\tilde{t})=\Lambda_h(1)+\left[a_h^r-\Lambda_h(1)\right]\phi_h^r\left(\frac{\tilde{t}-1}{-\varepsilon}\right),
    \end{equation}
    where $\phi_h^r(t)$ is defined as $\phi_h^r(t)=b_h^rt^2-(b_h^r-1)t^3$, $a_h^r$ and $b_h^r$ are undetermined coefficients. It is easy to verify that $\Lambda_h^{*r}(\tilde{t})$ satisfies the boundary conditions of $\Lambda_h$ and $\dif{\Lambda_h}/\dif{\tilde{t}}$ at $\tilde{t}=1$.\\
    To realize smooth connections between $\Lambda_h^{*l}$ and $\Lambda_h^{*m}$, the values of $a_h^l$ and $b_h^l$ should satisfy
    \begin{align}
        &\Lambda_h^{*l}(\tilde{t}=\varepsilon)=\Lambda_h^*(\tilde{t}=\varepsilon),\\
        &\left.\frac{\dif{\Lambda_h^{*l}}}{\dif{\tilde{t}}}\right|_{\tilde{t}=\varepsilon}=\left.\frac{\dif{\Lambda_h^*}}{\dif{\tilde{t}}}\right|_{\tilde{t}=\varepsilon}.
    \end{align}
    Similarly, $a_h^r$ and $b_h^r$ should satisfy
    \begin{align}
        &\Lambda_h^{*r}(\tilde{t}=1-\varepsilon)=\Lambda_h^*(\tilde{t}=1-\varepsilon),\\
        &\left.\frac{\dif{\Lambda_h^{*r}}}{\dif{\tilde{t}}}\right|_{\tilde{t}=1-\varepsilon}=\left.\frac{\dif{\Lambda_h^*}}{\dif{\tilde{t}}}\right|_{\tilde{t}=1-\varepsilon}.
    \end{align}
    By solving the above four equations, we obtain the explicit expressions of $\Lambda_{h}^{*l}(\tilde{t})$ and $\Lambda_{h}^{*r}(\tilde{t})$. In the limit of $\varepsilon\to 0$, $\Lambda_h^{*l}$, $\Lambda_h^{*m}$ and $\Lambda_h^{*r}$ lead to the same value of integral~(\ref{Ch-func}) with $\Lambda_h^*$ and hence they constitute the optimal $\Lambda_h(\tilde{t})$ that satisfy the boundary conditions of $\Lambda_h$ and $\dif{\Lambda_h}/\dif{\tilde{t}}$ at $\tilde{t}=0$ and $\tilde{t}=1$. The optimal $\Lambda_c(\tilde{t})$ can be determined in the same procedure. 

    \section{\label{appendix:B}Numerical method of determining the maximum power at given efficiency}
    To find the maximum power at given efficiency, we need to solve the equation $\partial \tilde{P}/\partial L_h=0$ for $L_h$. According to Eq.~(\ref{Ptilde-eta}), the equation turns out to be
    \begin{equation}
        \label{P-partial_Lh}
        (1+\delta-2L_h)L_h^2+b(\delta-L_h)^2(-1+2L_h)=0,
    \end{equation}
    where $b=(1-\eta)/[(1-\eta_{\C})\chi]$ and $\delta=(\eta_{C}-\eta)/(1-\eta)$. This is a cubic equation. It has three roots and what we need is the real roots $L_h^{*}$ that satisfy $0\le L_h^*\le\delta$. The analytical solutions are so cumbersome that it is hard to determine the proper solutions. Hence, we numerically solve Eq.~(\ref{P-partial_Lh}) at certain efficiency with given $\eta_{\C}$ and $\chi$ and pick out the proper solutions $L_h^*$. Then we compare the value of $\tilde{P}$ at $L_h=L_h^*$ with the boundary values ($\tilde{P}$ at $L_h=0,\delta$) and obtain the maximum value of power at certain efficiency with given $\eta_{\C}$ and $\chi$.
    
\end{document}